\documentclass[10pt, conference, letterpaper]{IEEEtran}
\IEEEoverridecommandlockouts
\usepackage{cite}
\usepackage{amsmath,amssymb,amsfonts}
\usepackage{graphicx}
\usepackage{textcomp}
\usepackage{xcolor}
\def\BibTeX{{\rm B\kern-.05em{\sc i\kern-.025em b}\kern-.08em
    T\kern-.1667em\lower.7ex\hbox{E}\kern-.125emX}}

\usepackage{bm}
\usepackage{mathtools, subfigure}
\usepackage{algorithm}
\usepackage{algpseudocode}
\usepackage{multicol}
\usepackage{subcaption}
\usepackage{authblk}

\newif\ifcomments

\commentstrue

\usepackage[absolute,showboxes]{textpos}
  
\TPMargin{5pt}
  
\newcommand{\copyrightstatement}{
    \begin{textblock}{0.84}(0.08,0.01)    
         \noindent
         \footnotesize
         \copyright 2023 IEEE. Personal use of this material is permitted. Permission from IEEE must be obtained for all other uses, in any current or future media, including reprinting/republishing this material for advertising or promotional purposes, creating new collective works, for resale or redistribution to servers or lists, or reuse of any copyrighted component of this work in other works.
    \end{textblock}
}

\begin{document}

\copyrightstatement
\title{Experimental Study of Transport Layer Protocols for  Wireless Networked Control Systems  \thanks{The authors acknowledge the financial support by the Federal Ministry of Education and Research of Germany in the programme of “Souverän. Digital. Vernetzt.”. Joint project 6G-life, project identification number: 16KISK002}
\thanks{The work of N. Pappas has been supported in part by the Swedish Research Council (VR), ELLIIT, and the European Union (ETHER, 101096526).}}

\author[*]{Polina Kutsevol}
\author[*]{Onur Ayan}
\author[**]{Nikolaos Pappas}
\author[*]{Wolfgang Kellerer}
\affil[*]{Chair of Communication Networks, Technical University of Munich, Germany}
\affil[**]{Department of Computer and Information Science, Link\"{o}ping University, Link\"{o}ping, Sweden}
\affil[ ]{Email: \{polina.kutsevol, onur.ayan\}@tum.de, nikolaos.pappas@liu.se, wolfgang.kellerer@tum.de}

\maketitle

\begin{abstract}

In Wireless Networked Control Systems (WNCSs), the feedback control loops are closed over a wireless communication network. The proliferation of WNCSs requires efficient network resource management mechanisms since the control performance is significantly affected by the impairments caused by network limitations. In conventional communication networks, the amount of transmitted data is one of the key performance indicators. In contrast, in WNCSs, the efficiency of the network is measured by its ability to facilitate control applications, and the data transmission rate should be limited to avoid network congestion. In this work, we consider an experimental setup where multiple control loops share a wireless communication network. Our testbed comprises up to five control loops that include Zolertia Re-Mote devices implementing IEEE 802.15.4 standard. We propose a novel relevance- and network-aware transport layer (TL) scheme for WNCSs. The proposed scheme admits the most important measurements for the control process into the network while taking current network conditions into account. Moreover, we propose a mechanism for the scheme parameters adaptation in dynamic scenarios with unknown network statistics. Unlike the conventional TL mechanisms failing to provide adequate control performance due to either congestion in the network or inefficient utilization of available resources, our method prevents network congestion while keeping the control performance high. We argue that relevance- and network-awareness are critical components of network protocol design to avoid control performance degradation in practice.
\end{abstract}

\section{Introduction}

\begin{figure}[t]
  \centering
  \includegraphics[width=0.8\linewidth]{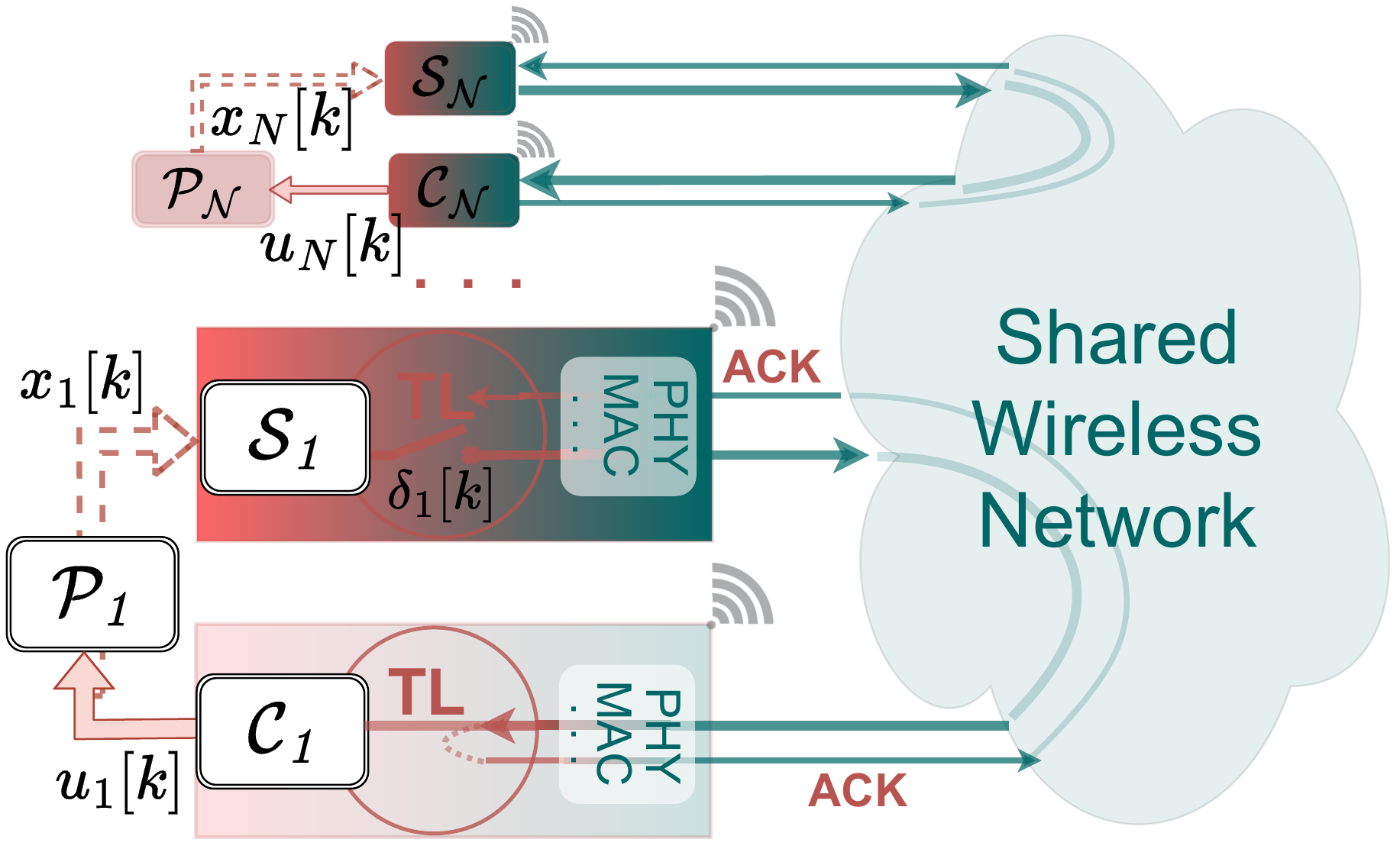}
  \caption{Considered scenario with $N$ control loops, where sensors' measurements are sent to corresponding controllers via a shared wireless communication network.}
  \label{fig:ncs_overview}
\vspace*{-5mm}
\end{figure}

Within Cyber-Physical Systems (CPSs), sensing, actuating and computing are integrated into feedback loops to control physical processes. Examples of CPSs relying on real-time control are smart homes, autonomous driving, surveillance systems and smart grids \cite{pivoto2021cyber}. The distributed deployment of control feedback loops enables novel use cases, where enhanced flexibility, mobility, cost efficiency and eased evolution are achieved if wireless networks are used for communication between the components\cite{dotoli2019overview}. The control loops where the feedback is closed over a wireless network are called Wireless Networked Control Systems (WNCSs). As shown in Fig.\ref{fig:ncs_overview}, the control loop includes the physical plant $\mathcal{P}_i$, the state of which is observed by the sensor $\mathcal{S}_i$. The sensor transmits status updates to the controller $\mathcal{C}_i$ over the network, which can be shared among multiple loops. The controller decides on the input to be applied to the plant to drive it to the desired state.

Along with certain benefits, wireless networks bring significant limitations, e.g., reduced reliability and non-negligible communication delays of sensors' measurements that affect the control performance. Indeed, if the controller does not receive fresh state updates, its control inputs can be sub-optimal. It potentially leads to the destabilization of the plant, jeopardizing the achievability of the control goal. Thus, it is important to develop network resource management techniques that consider the requirements of the control applications.

Current approaches to network resources management for WNCSs mostly disintegrate into two areas, namely system-theoretic and networking research, with a tenuous connection in between. In control theory, event-triggering (ET) \cite{molin2017event, balaghiinaloo2021decentralized} is often used to limit the volume of communicated data and offload the network. With ET, updates are transmitted only upon the occurrence of the event indicating the transmission relevance, e.g., state exceeding the threshold. Apart from ET, control-aware radio resources distribution between loops, e.g., scheduling minimizing control cost \cite{ma2022noisy} or estimation error \cite{kiekenap2020optimum}, is considered theory. However, these approaches usually rely on simplistic assumptions about the communication network, such as negligible or constant delays and packet loss, limiting their applicability in real scenarios. Meanwhile, conventional networking metrics such as throughput and delay can not capture the needs of control applications, and the optimizations w.r.t. these metrics are often detrimental to the control goal \cite{beytur2020towards}. To achieve high control performance in scenarios with limited network resource availability, communication algorithms should be both control- and network-aware.

The ideas of cross-layer design of network algorithms w.r.t. application goals utilizing application-defined metrics currently gain attention within the semantic communication paradigm \cite{kosta2017age, uysal2022semantic}. In semantic communication, it is proposed to introduce the metrics that consider the significance of the communicated data w.r.t. the application goal. Age of Information (AoI) \cite{kosta2017age} is a widely used semantic metric for real-time applications. AoI is the time elapsed since the generation of the last update available at the receiver. The minimization of AoI allows for finding the equilibrium between the under- and over-utilization of the network. Indeed, a very low sending rate of the updates results in high AoI due to rare actualization of the system state at the controller, while too frequent updates can congest the network, leading to increased AoI and degraded control performance.

Even though AoI goes beyond the conventional networking metric of delay, it does not exploit the relevance of the transmitted information. Several works \cite{mamduhi2020freshness, carabelli2017state} show that explicit consideration of the state dynamics and its importance for control goal leads to better control performance compared to the schemes based on AoI. Semantic metrics beyond age are shown to be effective at the medium access control (MAC) layer for wireless resource scheduling w.r.t. control goal both theoretically \cite{soleymani2021value, ma2022scheduling, nikkhah2023age} and in practice \cite{ayan2022task}. However, the amendment of lower MAC and physical (PHY) communication layers may not be favorable from the industry side because it narrows down the application area of devices. Due to cost limitations, real-life deployment scenarios are typically hardware-specific and come with hardware-induced limitations w.r.t. control performance. Thus, we focus on the transport layer (TL) design, which can be easily modified for the control performance improvement while being agnostic to the underlying hardware. \textit{Thereby, the flow injected into the network can be filtered at the TL based on the relevance of updates for the control goal, simultaneously avoiding network congestion.} For the TL schemes developed for the generic network, real-life performance validation is a vital issue. In this work, we deploy TL schemes on low-power low-cost IoT devices utilizing IEEE 802.15.4 standard \cite{7460875}. Namely, we use Zolertia Re-Mote sensor devices \cite{zolertia} operating in 2.4 GHz unlicensed spectrum. Zolertia sensors are ready to be deployed in industrial scenarios and represent a resilient solution for smart cities, e-health, etc.

\vspace*{-2mm}
\subsection{Contributions}
The main contributions of this work are the following:
\begin{itemize}
    \item  We propose an ET scheme for TL protocol for WNCSs where the triggering rule considers both the relevance of the next status update for the control goal and the current network congestion level. To assess the relevance of updates, we develop a novel scheme for the augmentation of the controller state at the sensor that is applicable in practical scenarios with non-perfect network conditions.
    \item We design a distributed threshold adaptation (TA) scheme for the scenarios where the optimal threshold value for ET can not be set in advance, such as when sensors are unaware of the network conditions or when the network state is varying over time. The proposed scheme is tailored for real-life deployment, as it takes into account the potential variability in the wireless channel quality of different devices. To our knowledge, no previous studies are developing TA schemes for WNCSs depending on the network state and the individual channel qualities of devices.
    \item We study the performance of proposed schemes with Zolertia Re-Mote sensors in scenarios with up to 5 control loops sharing the wireless network. Thus, we present the first work where relevance-aware TL schemes for WNCSs are implemented on real hardware. Our results show that the suggested TL mechanism outperforms conventional and state-of-the-art  (SotA) schemes from both system theoretic and networking domains, revealing the potential of control-aware network protocol design. The enhanced ET scheme that uses the controller state augmentation improves the control performance by at least $10\%$ in scenarios with intensive network traffic.
    \item We show the effectiveness of the TA scheme in scenarios where the number of control loops is unknown or varies over time. The TA mechanism shows more than $40\%$ performance improvement compared to the fixed-threshold mode, whereas the enhanced ET scheme results in a further $30\%$ decrease in control costs. Finally, experiments with different MAC layer configurations demonstrate the versatility of the proposed ET schemes and the TA mechanism.
    \end{itemize}

\section{Related Work}

\label{sec:related_work}

For WNCSs, it is critical to appropriately determine the sending rate of the updates at a TL, because, as shown in \cite{gautam2021comprehensive}, control performance degrades for both too low and too high sending rates. As shown in \cite{beytur2020towards, shireliable}, conventional networking TL policies are inefficient for real-time applications under the increased traffic load. For instance, increased delays and drop rates are peculiar to Transmission Control Protocol (TCP) and User Datagram Protocol (UDP) protocols, respectively, which deteriorates the control performance.

In control theory, the prevalent method to avoid network congestion is ET. ET leads to a significant reduction of the network traffic while keeping adequate control performance. Motivated by potential communication savings, the works \cite{molin2017event, funk2021learning} discuss triggering policies based on the state itself and \cite{ramesh2016performance, balaghiinaloo2021decentralized} - based on the difference between the state and its estimation by the controller. To our knowledge, no ET mechanisms that adapt the threshold value to instantaneous network conditions have been developed prior this work. The authors of \cite{carabelli2017state, kiekenap2020optimum, ma2022optimal} discuss centralized radio resource scheduling approaches to avoid congestion-related effects on control performance.

With the proliferation of semantic communications \cite{kountouris2021semantics, uysal2022semantic, strinati20216g, wheeler2022engineering, kosta2017age}, many works introduce new semantic metrics for effective network management w.r.t. control goal. \cite{beytur2020towards} analyses AoI for real-life connections utilizing TCP and UDP. The authors confirm that AoI can capture the control performance degradation if the network is under-utilized or congested. Age Control Protocol (ACP) for TL designed in \cite{shreedhar2019age} adapts the sending rate to minimize AoI. Age-based metrics are used at the MAC layer for scheduling in \cite{ouguz2022implementation, mamduhi2020freshness}. In particular, the authors of \cite{ouguz2022implementation} state that link reliability should be considered together with AoI when prioritizing control loops. The work \cite{mamduhi2020freshness} shows that triggering the transmission of updates based on the state estimation error in combination with AoI-minimizing scheduling is more beneficial for control costs than AoI-based resource management. Indeed, freshness maximization does not always result in the best control performance because AoI does not explicitly consider the process dynamics. This notion has given a rise to metrics beyond age, such as Value of Information \cite{soleymani2021value, ayan2019age}, Age of Incorrect Information \cite{maatouk2020age}, Deviation of Information \cite{noroozi2022age}, Age of Actuation \cite{nikkhah2023age}. These works discuss the benefits of plugging relevance-aware measures into network management policies design, such as reduced network utilization, improved control cost or decreased estimation error. 

The majority of the discussed works validate their approaches with simulations and assume networking delays below one sampling period \cite{ramesh2016performance, balaghiinaloo2021decentralized, kiekenap2020optimum, soleymani2021value, ayan2019age, maatouk2020age} or constant \cite{carabelli2017state}. The packet loss is often considered to be constant \cite{maatouk2020age} or negligible \cite{kiekenap2020optimum, mamduhi2020freshness, soleymani2021value, ayan2019age, noroozi2022age}. Many works assume instant knowledge of the estimator state at the receiver \cite{ramesh2016performance, balaghiinaloo2021decentralized, mamduhi2020freshness} or even a perfect channel \cite{molin2017event}. Very few experimental works with real hardware exist, but they also consider scenarios with the networking delay under one communication slot \cite{ma2022optimal, ouguz2022implementation}, perfect acknowledgments \cite{ouguz2022implementation} or do not provide details on networking setup \cite{funk2021learning}.

Real-life experiments with multiple control loops utilizing different MAC layer policies are conducted in \cite{ayan2022task}. This work shows that control- and network-aware scheduling is more effective than age-based mechanisms. The authors of \cite{shreedhar2019age} implement the TL mechanism for WNCSs, without making assumptions about the underlying network. Even though ACP proposed in \cite{shreedhar2019age} is shown to be efficient in terms of AoI minimization, the authors do not discuss the performance of end applications and semantic metrics beyond age. Thus, our work is the first one addressing the potential of relevance-awareness for TL mechanisms for WNCSs in practice.
\section{System Model}
\label{sec:system_model}
We consider a wireless network shared by $N$ linear time-invariant (LTI) control systems, as shown in Fig.\ref{fig:ncs_overview}. Each system $i \in \{1, \dots, N\}$ contains a plant $\mathcal{P}_i$, the state of which is periodically observed by a sensor $\mathcal{S}_i$. Controllers $\mathcal{C}_i$ are co-located with plants $\mathcal{S}_i$, whereas sensors operate remotely. We assume that sensors perfectly capture the plant states. While the plant-to-sensor and controller-to-plant links are assumed to be ideal, sensors transmit the measurements to controllers via a non-ideal wireless link in the form of status update packets, with each packet carrying single measurement that corresponds to the payload of $20$B\footnote{As shown in Fig.~\ref{fig:rtts}, in our setup, the concatenation of multiple measurements in one packet leads to an RTT increase. Thus, the benefits of the concatenation are questionable, and we do not consider it in this work.}.

Upon the generation, each status update packet is either admitted to the network or discarded based on some criteria. The selected TL strategy defines the admission policy. The access to the shared medium is independent of the TL, and the TL does not have further control over already admitted packets. The operation of the TL does not rely on link layer-specific information but only assumes an end-to-end acknowledgment (ACK) procedure between each sensor and controller. Each packet is subject to networking delay and potential loss. The updates successfully received by the controller refine its knowledge on the plant state. Based on this information, the controller decides on further control inputs. 
\subsection{Control Model}
The dynamics of each LTI system are modeled in the discrete-time as follows:
\begin{equation}
	\bm{x}_i[k+1] = \bm{A}_i\bm{x}_{i}[k] + \bm{B}_i\bm{u}_i[k] + \bm{w}_i[k],
	\label{eq:dynamics}
\end{equation}

where $\bm{x}_i[k] \in \mathbb{R}^{n_i}$ is the state of the plant $\mathcal{P}_i$ at the time step $k$. The sensor and the controller within each loop are synchronized in time.  $\bm{u}_i[k] \in \mathbb{R}^{m_i}$ defines a control input calculated by the controller $\mathcal{C}_i$ at the time step $k$.  $\bm{A}_i \in \mathbb{R}^{{n_i}\times {n_i}}$ and $\bm{B}_i \in \mathbb{R}^{{n_i}\times {m_i}}$ are time-invariant state and input matrices. These matrices do not vary over the considered time horizon, and their values are known to the controller. The component introducing the randomness into the plant state is the process noise, the value of which at the time step $k$ is denoted with $\bm{w}_i[k] \in \mathbb{R}^{n_i}$. The sampled noise vectors are independent and identically distributed (i.i.d) according to a zero-mean Gaussian distribution with a covariance matrix $\bm{W}_i \in \mathbb{R}^{{n_i}_ \times {n_i}}$.

Based on the available measurements, the controller estimates the current system state and specifies the control input. The goal of the controller is to minimize the control cost. In this work, we use the quadratic control cost function applicable for linear systems with Gaussian noise referred to as linear-quadratic Gaussian (LQG) cost. LQG cost includes the state deviation from the desired point and the control effort spent to keep the plant stable:
\begin{equation}
	\mathcal{J}_i \triangleq \limsup _{\mathcal{T}\rightarrow \infty}  \left(\dfrac{1}{\mathcal{T}} \sum_{k=0}^{\mathcal{T} - 1} (\boldsymbol{x}_i[k])^T \boldsymbol{Q}_i \boldsymbol{x}_i[k] +  (\boldsymbol{u}_i[k])^T \boldsymbol{R}_i \boldsymbol{u}_i[k] \right),
	\label{eq:lqg}
\end{equation} 
where $\mathcal{T}$ is the time horizon, over which the system is considered,  $\boldsymbol{Q}_i$ and $\boldsymbol{R}_i$ regulate the prioritization of control effort minimization over deviation or vise versa. $\bm{x}_i^T[k]$ and $\bm{u}_i^T[k]$ are obtained from  $\bm{x}_i[k]$ and $\bm{u}_i[k]$ by transposition.  Note that the desired state of the systems is fixed to $\bm{0} \in \mathbb{R}^{n_i}, \forall i$.

The controller minimizing the LQG cost in \eqref{eq:lqg} follows the linear-quadratic regulator design:
\begin{equation}
	\label{eq:controllaw}
	\bm{u}_i[k] = - \bm{K}_i \bm{\hat{x}}_i[k],
\end{equation}

where $\bm{K}_i$ is the optimal feedback matrix and $\bm{\hat{x}}_i[k]$ is the state estimation at the controller based on the available plant state measurements. According to the separation principle known in the control theory, if the process observations are Gaussian, optimal estimation and feedback rule can be designed independently. Thus, we assume that $\bm{K}_i$ is designed before the system deployment, i.e., independent of the underlying network, and does not have to be re-designed if the network conditions change. $\bm{K}_i$ can be found as:
\begin{equation}
	\label{eq:K_i}
	\bm{K}_i = (\bm{R}_i + \bm{B}_i^T\bm{P}_i\bm{B}_i)^{-1}\bm{B}_i^T\bm{P}_i\bm{A}_i,
 \end{equation}
where $\bm{P}_i$ is a solution of the discrete-time algebraic Ricatti equation. Here, $\bm{(\cdot)}^{-1}$ indicates a matrix inversion.

Now, let us consider the design of the state estimation procedure at the controller. Let $\nu_i(k)$ be the sampling time step of the freshest update at the controller. Following the Kalman filter design, similar to \cite{ayan2022task}, the state estimation is found as an expectation of the system state given the observation $\bm{x}_i[\nu_i(k)]$:
\begin{equation}  
	\bm{\hat{x}}_i[k] = \bm{A}_i^{\Delta_i[k]} {\bm{x}}_i[\nu_i(k)] + \sum_{q =1}^{\Delta_i[k]} \bm{A}_i^{q - 1} \bm{B}_i \bm{u}_i[k - q],
	\label{eq:estimator}
\end{equation}
with $\Delta_i[k] = k - \nu_i(k) $ being the AoI at the controller in the units of a number of sampling periods. Here, matrix $\bm{M}^l$ is $l$-th power of matrix $\bm{M}$. Note that the exact reproduction of the system state at the controller based on the delayed measurement is impossible due to the presence of the disturbances $\bm{w}_i$. The estimation error is represented as:
\begin{equation}
	\bm{e}_i[k] = \bm{x}_i[k] - \bm{\hat{x}}_i[k] = \sum_{q =1}^{\Delta_i[k]} \bm{A}_i^{q-1}\bm{w}_i[k-q].
	\label{eq:est_error}
\end{equation}
As one can conclude from \eqref{eq:est_error}, the instantaneous estimation error depends on the current AoI rather than the networking delay. \textit{This motivates the consideration of age as a semantic metric that should be minimized for efficient radio resource management for control applications.} However, the minimization of AoI does not imply the minimization of either the average estimation error or the control cost from \eqref{eq:lqg}. The estimation error \eqref{eq:est_error} is not determined by the instantaneous AoI, as it depends on the random disturbance and thus on the exact dynamics of the state. That is why, the consideration of the current state $\bm{x}_i[k]$ is potentially more beneficial for extracting the relevance of a particular update for the control process than the AoI.

\subsection{Network Model}

At TL, each sensor makes a decision $\delta_i[k] \in \{0,1\}$ regarding the admission of a current measurement to the network based on the state dynamics and the information extracted from the TL ACKs. TL ACKs include the generation and the reception time steps of the updates. $\delta_i[k] \in \{0,1\}$ is a binary decision variable and equals to $1$ if the update $\bm{x}_{i}[k]$ is admitted to the network. Note that if not admitted, the sampled update is discarded without further consideration. For acknowledged packets, the sensor keeps round-trip time (RTT) statistics, which is a time duration between the moments the update $\bm{x}_{i}[k]$ is admitted to the network and ACK for this update is received. 

Our goal is to design a TL policy $\bm{\pi^*}$, such that the average control cost \eqref{eq:lqg} is minimized:
\vspace*{-4mm}
\begin{equation}
    \label{eq:opt_prob}
    \bm{\pi^*} = \arg \min_{\bm{\pi}} \frac{1}{N} \sum_{i=1}^N \mathcal{J}_i.
\end{equation}
Generally, a good TL policy should be able to identify the maximum average injection rate of packets to the lower layers such that the network congestion level does not degrade the control performance. Furthermore, we take advantage of the state dynamics and ACKs available to the sensor to prioritize the most relevant updates and to track instantaneous network conditions. Indeed, based on RTTs derived from ACKs, the sensor can monitor the current network state and the congestion level. Moreover, ACKs provide the sensor with information on which updates are available at the controller and when the controller has received them. The sensor can use this information to augment the controller estimation process and refine the relevance of the current update for this process. 
\section{Transport Layer Protocols for WNCS}
\label{sec:policies}

\subsection{Conventional and SotA Transport Layer Schemes}
We implement and analyze conventional networking TL policies such as UDP, TCP and Zero-Wait (ZW).  Also, we consider an ET scheme that does not take the network state into and age-minimizing ACP from \cite{shreedhar2019age}.
\subsubsection{UDP}
According to UDP, all the packets arriving at the TL are admitted to the network. The policy follows the rule 
\begin{equation}
    \label{eq:udp}
    \delta_i^{UDP}[k] = 1, \forall i, \forall k.
\end{equation}
UDP neither requires connection establishment nor tracks if the transmitted packets arrive at the destination node. Thus, UDP is a fast but unreliable scheme, agnostic to the network congestion level. Thus, UDP usage is expected to result in poor control performance in scenarios with constrained network resources.
\subsubsection{TCP}
If requirements on the reliability of the data delivery are prevalent, TCP is often used instead of UDP. TCP guarantees that all the transmitted data reaches the destination while maximizing the throughput. For that, TCP utilizes a delivery ACK scheme. TCP tracks the amount of currently unacknowledged packets called outstanding packets (OPs) and does not admit new updates to the network if the number $n^{out}_i[k]$ of OPs is higher than the specified congestion window value $CW_i[k]$. An update is added to the OP list when pushed to the MAC layer and deleted when ACK for this packet is received or if the time greater than ACK timeout has elapsed since its transmission. Note that this can happen due to the lost update or ACK packet since we do not assume the ACK link is error-free. 

Reliability in TCP is achieved through retransmissions. Namely, if a packet is considered lost, the same content is pushed to the network again. However, in the WNCSs scenario, by the time the update is known to be lost, new measurements are sampled by the sensor, representing fresher information. Since only the freshest update is relevant for the controller (see \eqref{eq:estimator}) retransmitting outdated packets is needless. Conversely, retransmissions are detrimental to control since they congest the network with stale packets. Thereby, for our analysis, we adopt the congestion window adaptation of TCP and do not allow retransmissions, stating that the control performance of the considered scheme is not worse than that of the original TCP. The admission rule for TCP is the following:   
\begin{equation}
    \label{eq:tcp}
    \delta_i^{TCP}[k] = 
    \begin{dcases}
    1,& \text{if } n^{out}_i[k] < CW_i[k]\\
    0,              & \text{otherwise.}
    \end{dcases}
\end{equation}

 The $CW_i[k]$ adaptively changes according to the congestion control scheme to maximize throughput. In this work, we study Tahoe and Vegas congestion control algorithms. Congestion control of TCP Tahoe considers packet drop events, whereas TCP Vegas is delay-based.
 
\subsubsection{Zero-Wait}
Similarly to TCP, the ZW policy \cite{sun2019sampling} utilizes ACKs. However, here $CW_i[k]$ is set to $1$. The ZW admission policy reads as 
\vspace*{-3mm}
\begin{equation}
    \label{eq:zw}
    \delta_i^{ZW}[k] = 
    \begin{dcases}
    1,& \text{if } n^{out}_i[k] = 0 \\
    0,              & \text{otherwise.}
    \end{dcases}
\end{equation}

Thus, ZW represents a ``stop-and-wait" policy where after sending each update, the receiver waits for it to be acknowledged or identified as lost before sending the next one. Such a technique can prevent packets' queuing in the MAC layer if the MAC buffer utilizes a first-in-first-out (FIFO) strategy. Instead of becoming outdated while waiting in the queue, the fresh updates are inserted into the empty buffer and thus can be delivered faster. 

\subsubsection{Conventional Event-Triggering}
This scheme considers the current state of the plant. In particular, 
\begin{equation}
	    \delta_i^{ET}[k] = \begin{dcases}
        1,& \text{if } |\bm{x}_i[k]| \ge \lambda_i\\
        0,              & \text{otherwise.}
    \end{dcases}
\end{equation}
ET admits to the network only relevant updates, indicating that the deviation from zero is higher than the threshold $\lambda_i$, thus offloading the network. Nevertheless, it can result in severe network congestion. Indeed, as soon as the state grows above the threshold, the ET starts pushing updates in a bursty fashion until the controller drives the plant back to zero, congesting the network. Since the scheme is agnostic to the current network state, it can not capture the situations when the transmission cost outweighs the benefits of transmitting an update. 
\subsubsection{Age Control Protocol}

The ACP policy is a TL protocol designed for real-time monitoring systems. It represents an iterative algorithm for adapting the sending rate to minimize the estimated age at the receiver. For the details on the implementation of ACP, we refer the reader to \cite{shreedhar2019age}.

 \subsection{Relevance- and Congestion-aware Event-Triggering Scheme for WNCSs}
For our proposed TL policy, we consider both the relevance of the transmitted data and the current network congestion level in terms of the number of OPs. Below, we present two schemes. The Zero-Wait Event Triggering (ZW ET) scheme uses the state deviation from zero, whereas the Augmentation-based Zero-Wait Event Triggering (AUGM ZW ET) mechanism considers the difference between the current state and its controller estimation. Note that the proposed policies are implemented on top of UDP, i.e., controller-sensor pairs establish a UDP connection and use UDP to transport updates and ACKs over the network. 
\subsubsection{Zero-Wait Event Triggering}
The deviation of the current state far from the set point typically indicates the necessity of the transmission, so the controller can drive the plant back to zero,  contributing to the reduction of the overall control cost. Simultaneously, similarly to ZW policy, in ZW ET, the sensor does not allow more than one OP in the network to prevent congestion. Thus, ZW ET considers the transmission cost in line with the relevance of the updates for the control process. ZW ET admits the measurements according to the following rule:
\vspace*{-3mm}
\begin{equation}
	   \delta_i^{ZW\_ET}[k] = \begin{dcases}
    1,& \text{if } |\bm{x}_i[k]| \ge \lambda^i  $\text{and} $ n_{out}^i[k] = 0\\
    0,              & \text{otherwise.}
\vspace*{-3mm}
\end{dcases}
\end{equation}

Below we highlight the fundamental difference between this approach and the mechanisms using AoI as a metric. The update can carry information about low deviation, even though the age recorded on the monitor is high. In this case, the update is not urgent, and its transmission congests the network while bringing minor benefits. \textit{ Thereby, the exploitation of the updates' content allows the extraction of more precise knowledge regarding the relevance of the measurement.}
\subsubsection{Augmentation-based Zero-Wait Event Triggering}

\begin{figure}[t]
    \centering
    \subfigure[Minimum RTT as a function of a payload size.]{\includegraphics[width=0.3\linewidth]
    {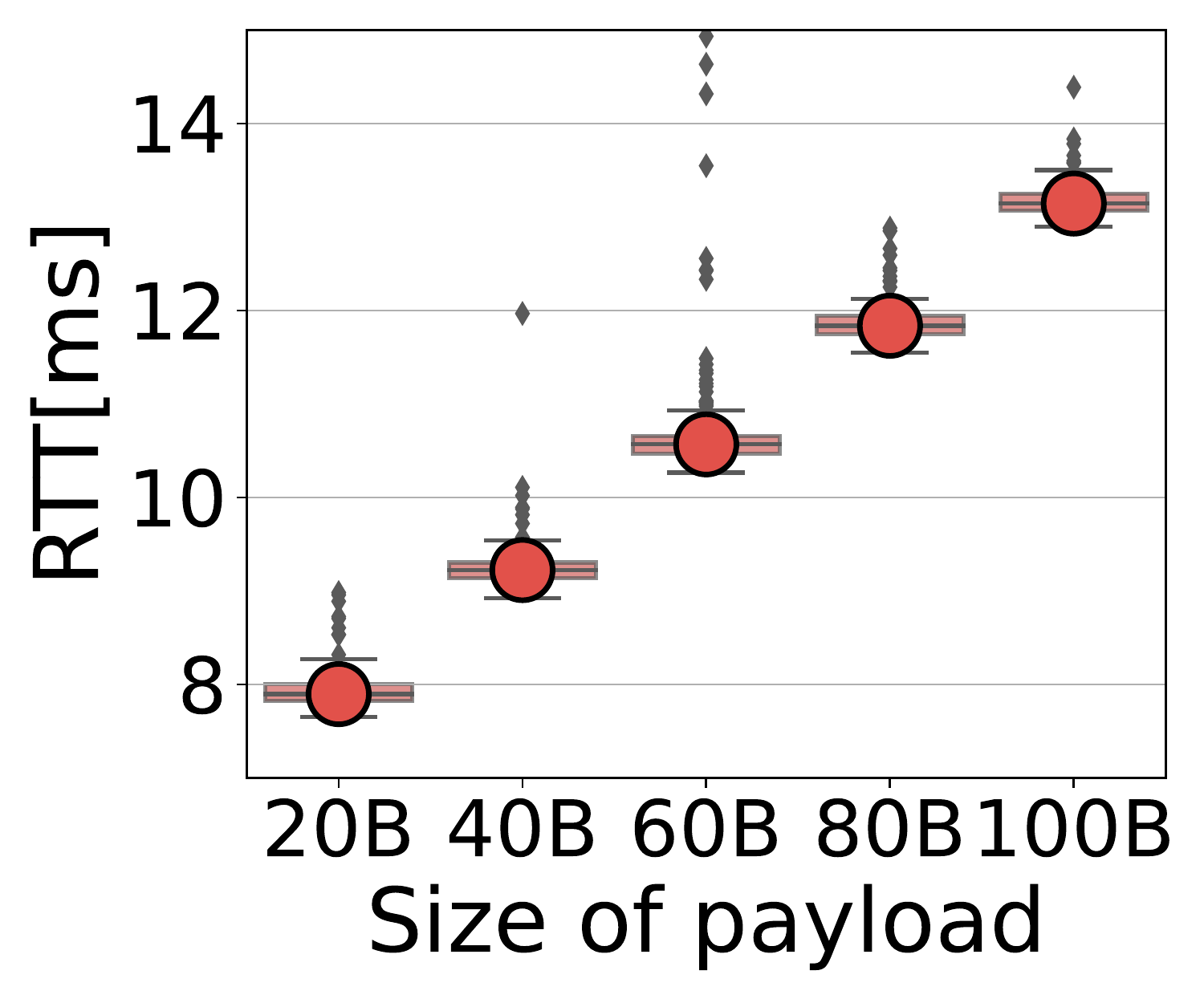} \label{fig:rtts}}
    \subfigure[RTT histogram for a loop with better channel conditions]{\includegraphics[width=0.3\linewidth]{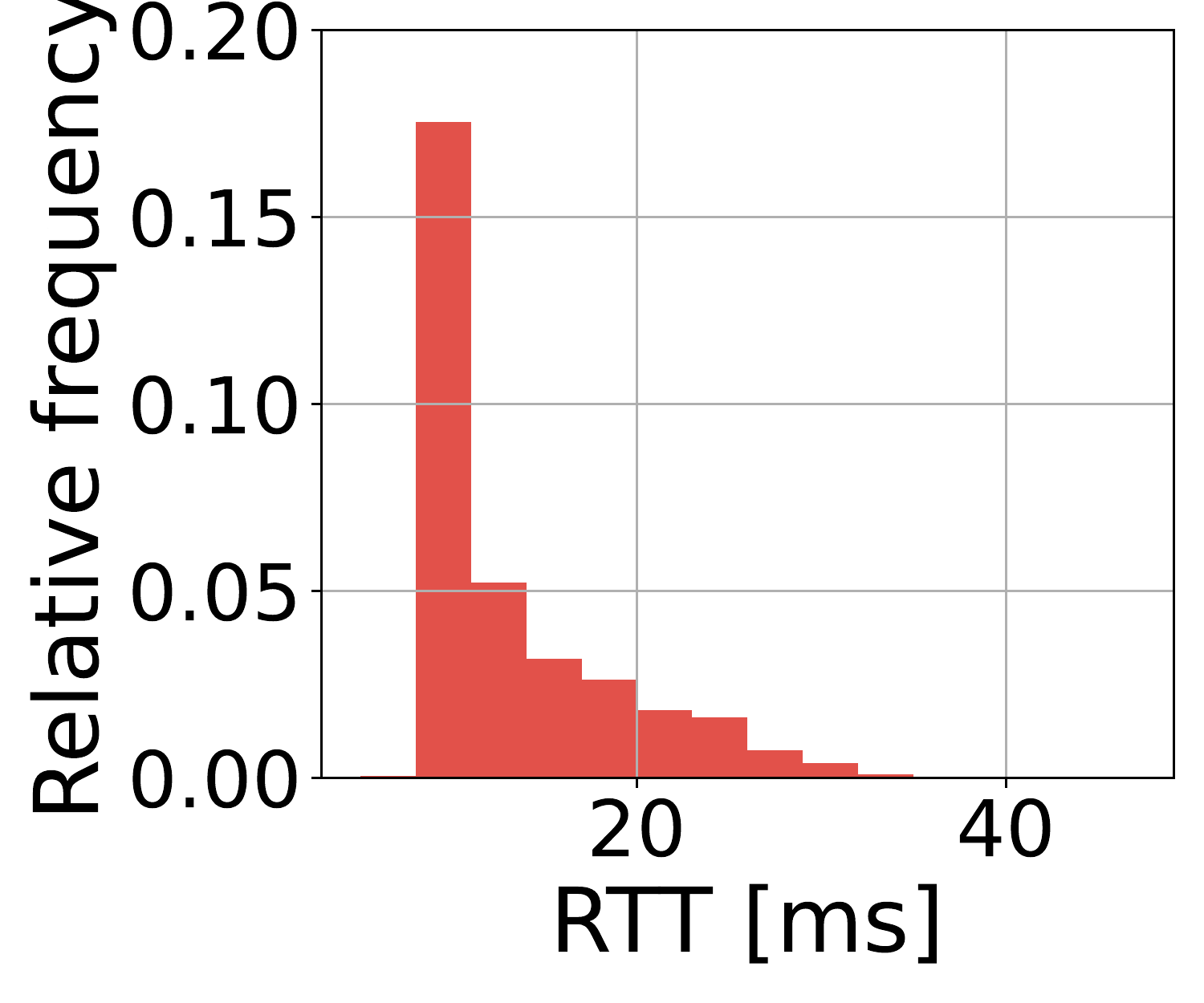}\label{fig:rtt_good}}
    \subfigure[RTT histogram for a loop with worse channel conditions]{\includegraphics[width=0.3\linewidth]{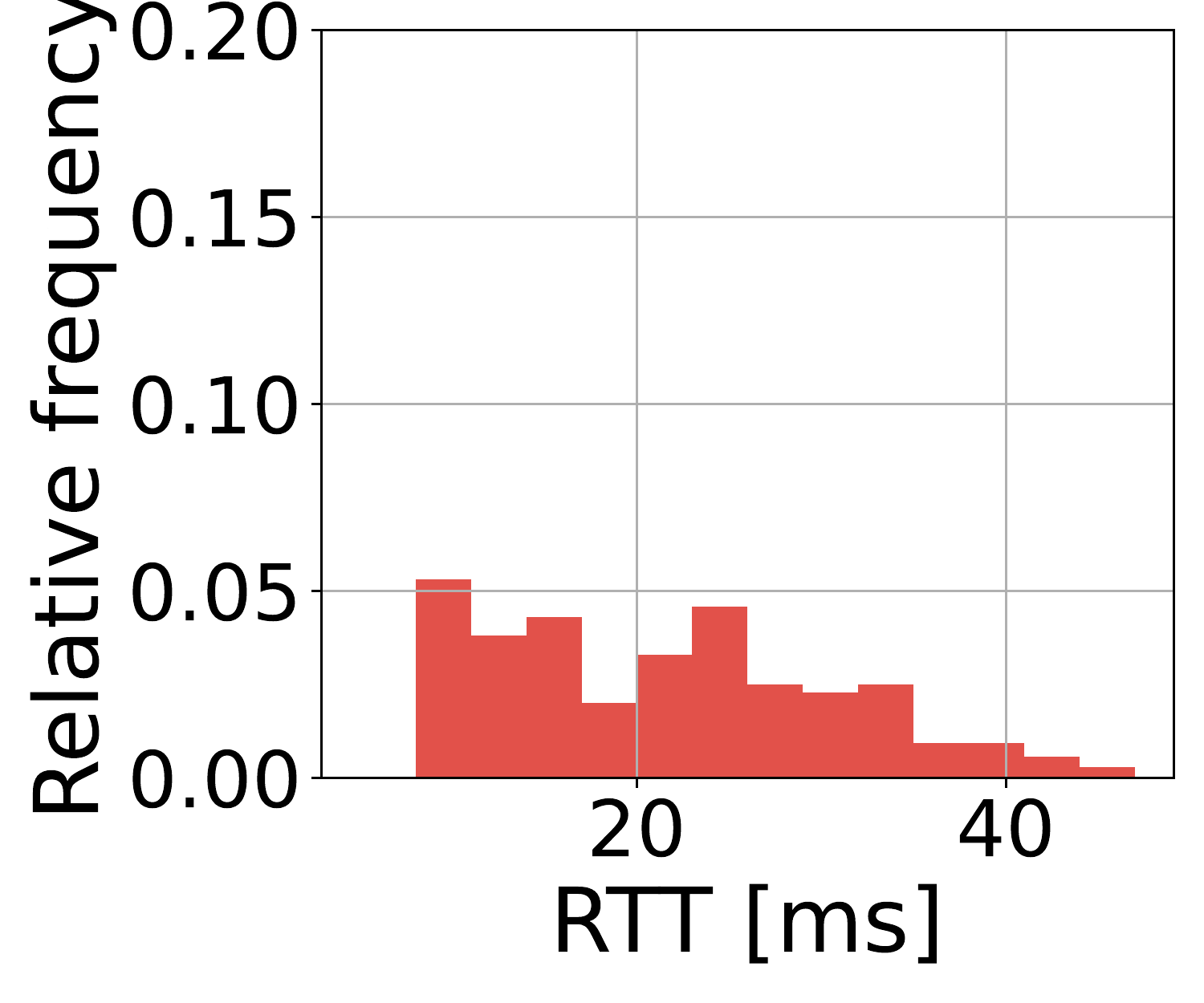} \label{fig:rtt_bad}}

\caption{Experimental round trip time statistics.}
\label{fig:rtt_hist}
 \vspace*{-5mm}
\end{figure}

Since the controller uses the state estimation to determine control inputs, the estimation error from \eqref{eq:est_error} is more informative than the magnitude of state w.r.t. the relevance of the update. Indeed, when the state only is considered for admission, some unnecessary transmissions are triggered when the plant deviates from zero, and the controller is already aware of it.

To avoid this unnecessary traffic, we propose to use ACK information at the sensor to augment the controller estimation. In particular, the sensor repeats the estimation process of the controller, assuming that the last ACKed packet coincides with the last received packet at the controller. Note that the idea of copying the controller process at the sensor side for assessing the controller estimation was proposed in the literature before, e.g., \cite{ramesh2016performance}. However, existing works assume a perfect ACK link, making the augmentation task trivial. Let $\bar{\nu}_i(k)$ be the sampling instance of the last ACKed update. Then, analogously to \eqref{eq:estimator}, the augmented controller estimation at the sensor can be expressed as follows:
\vspace*{-2mm}
\begin{equation}
	\bm{\bar{x}}_i[k] = \bm{A}_i^{k - \bar{\nu}_i(k)} \bm{x}_i[\bar{\nu}_i(k)] + \sum_{q =1}^{k - \bar{\nu}_i(k)} \bm{A}_i^{q - 1} \bm{B}_i \bar{\bm{u}}_i[k - q],
	\label{eq:augm_estimator}
 \vspace*{-2mm}
\end{equation}
where $\bar{\bm{u}}_i$ is control input augmented by a sensor. The sensor finds it by updating the augmented estimation \eqref{eq:augm_estimator} at each time step and calculating the control input similarly to \eqref{eq:controllaw}. Note that upon the reception of the new ACK, the sensor updates past inputs $\bar{\bm{u}}_i$ because it refines past augmented estimations. The admission rule for the AUGM ZW ET is:
\begin{equation}
	   \delta_i^{AUGM\_ZW\_ET}[k] = \begin{dcases}
    \begin{multlined}
       1, \;\;\;\;\text{if } |\bm{x}_i[k] - \bar{\bm{x}}_i[k]| \ge \lambda^i \\
        $\text{and} $ n_{out}^i[k] = 0 
    \end{multlined}\\
    0,  \;\;\;\; \text{otherwise.}
\end{dcases}
\end{equation}

\begin{algorithm}[t]
\caption{Threshold Adaptation}\label{alg:cap}
\footnotesize
\begin{algorithmic}
    \While{$\mathit{true}$}
        \If{$len(tmp\_rtt\_list) \ge N_b$}
            \If{$mean(tmp\_rtt\_list) > mean + f(stddev)$}
                \State $INC(\lambda)$
               
            \Else
                \State $DEC(\lambda)$
            \EndIf
                
            \State $mean \gets mean(rtt\_list[end - w: end])$
            \State $stddev \gets stddev(rtt\_list[end - w: end])$
            \State $tmp\_rtt\_list \gets NULL$
        \EndIf
    \EndWhile
\end{algorithmic}
\end{algorithm}
\vspace*{-3mm}
 \subsection{Threshold Adaptation}
The congestion in the network and, consequently, control performance strongly depends on the traffic intensity, which, in turn, is tightly connected to the threshold values $\lambda_i$. In this subsection, we present the novel Threshold Adaptation (TA) technique allowing dynamic adaptation of the traffic injection rates according to instantaneous network conditions while keeping the control performance high. It is vital in scenarios where the appropriate $\lambda_i$ can not be set in advance. Our TA mechanism is a distributed scheme at TL used by sensors to independently adjust $\lambda_i$ to a minimum value not causing severe network congestion and control performance degradation. 

We use the fact that the increase in RTTs observed by a sensor witnesses the increased congestion level in the network. The initial $\lambda_i$ should be high enough such that RTTs are close to the minimum value. The TA Algorithm \ref{alg:cap} is triggered when another batch of $N_b$ packets is ACKed. Observed mean RTT of these $N_b$ packets is compared to $mean + f(stddev)$, where $mean$ and $stddev$ are average statistics recorded for $w$ last RTTs. If the observed mean is higher than $mean + f(stddev)$, $\lambda_i$ is increased, otherwise - decreased. One should choose $w \gg N_b$ to capture average statistics. The margin $f(stddev)$ that depends on the standard deviation of RTT distribution allows capturing the difference in the link quality observed by different sensors. As supported by Fig.\ref{fig:rtt_good} and Fig.\ref{fig:rtt_bad} representing RTT statistics of two loops operating simultaneously, some sensors experience better channel conditions and have a more peaky RTT distribution (as in Fig.\ref{fig:rtt_good}) and smaller $stddev$ than others (as in Fig.\ref{fig:rtt_bad}). According to our algorithm, loops with smaller $stddev$ are less affected by congestion, and even a slight growth in average RTT should trigger $\lambda_i$ increase. At the same time, larger average RTT deviations are tolerated for loops with higher $stddev$. Such behaviour prevents too conservative $\lambda_i$ for the sensors with the worse channel, and a very high sending rate and increased resource consumption of the loops with the better link quality. 
 
 Note that instead of $f(stddev)$, one could use the margin that depends on the quantiles of the RTT distribution. However, our experiments revealed the inefficiency of such an approach since loops with different channel qualities often had similar tails of RTT distribution, making TA similarly sensitive to the RTT increase. Such behaviour can be explained by the fact that in our experiments, RTT is bounded by the maximum number of retransmissions at the MAC layer.

\section{Experimental Framework}
\label{sec:exp_framework}

In this section, we provide details on the hardware used for experiments and the general WNCS setup. We use Zolertia Re-Mote sensors following the IEEE 802.15.4 standard \cite{7460875}. The reason behind this choice of hardware is that the design of the developed TL scheme aims at control performance gains in real scenarios with real-life devices used for the deployment of emerging applications for CPSs. Zolertia Re-Mote is a resilient IoT platform ready to be deployed in industrial applications, Smart Cities, etc., i.e., in the networked control field, a target use case for the mechanisms discussed in the current work. Combined with our TL policy for WNCSs, Re-Mote sensors represent a fast solution for the smooth integration of control-based services. Thus, the advances of the proposed schemes with Zolertia hardware prove their applicability in typical IoT scenarios.

\subsection{Zolertia Re-Mote Architecture}

Zolertia Re-Motes are low-power low-cost wireless sensor nodes with built-in 2.4-GHz IEEE 802.15.4-compliant radio transceiver and Universal Asynchronous Receiver-Transmitter (UART) interface. Nodes use the Contiki-NG operating system \cite{1367266} for resource-constrained IoT devices, which implements the networking stack. For PHY, Contiki-NG implements IEEE 802.15.4 designed for low-cost communication, which supports a data rate of 250 kbps. MAC buffer is a FIFO queue. For channel access at the MAC layer, we use contention-based IEEE 802.15.4 Carrier Sense Multiple Access (CSMA) or a custom polling-based scheduling. The gateway node polls all the sensors in a Round-Robin fashion and grants the slot for the transmission if the sensor indicates that its MAC buffer is not empty. Further, sensors use CSMA unless otherwise stated.

\vspace*{-2mm}
\subsection{Measurement Setup}

\begin{figure}[t!]
    \centering
    
    \subfigure[Schematic representation]
    {\includegraphics[width=1\linewidth]{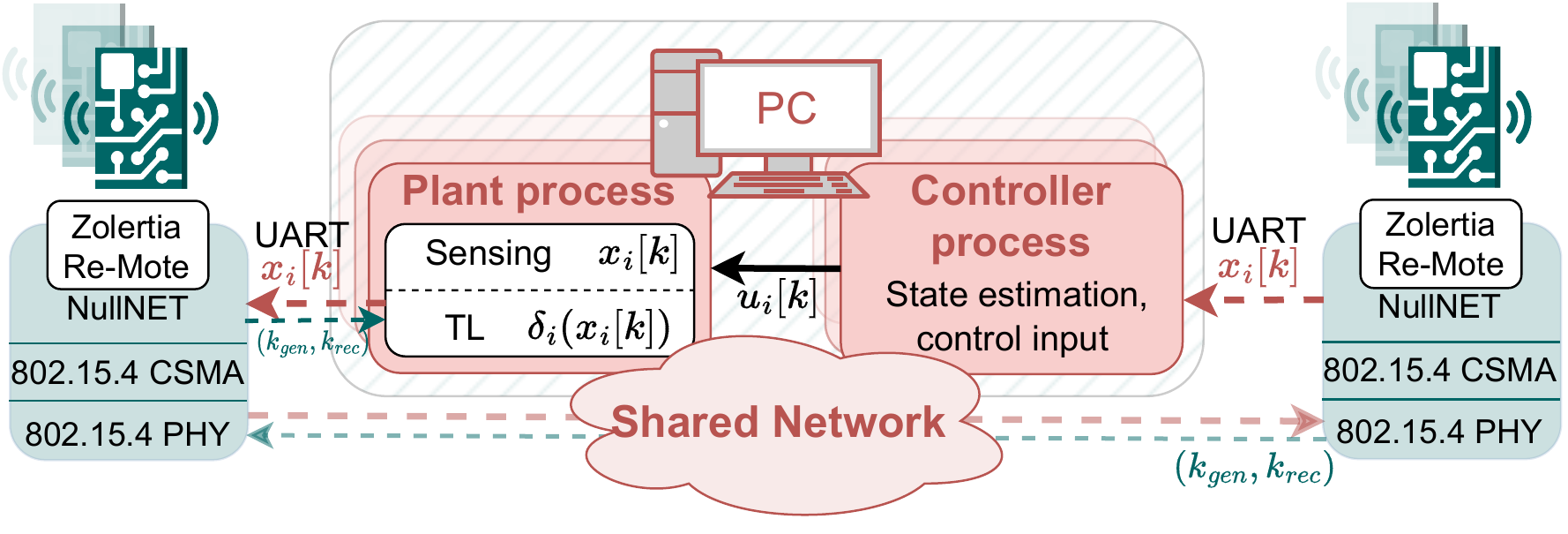}\par
    \label{fig:setup}}
    \subfigure[Experimental setup]{\includegraphics[width=0.7\linewidth]{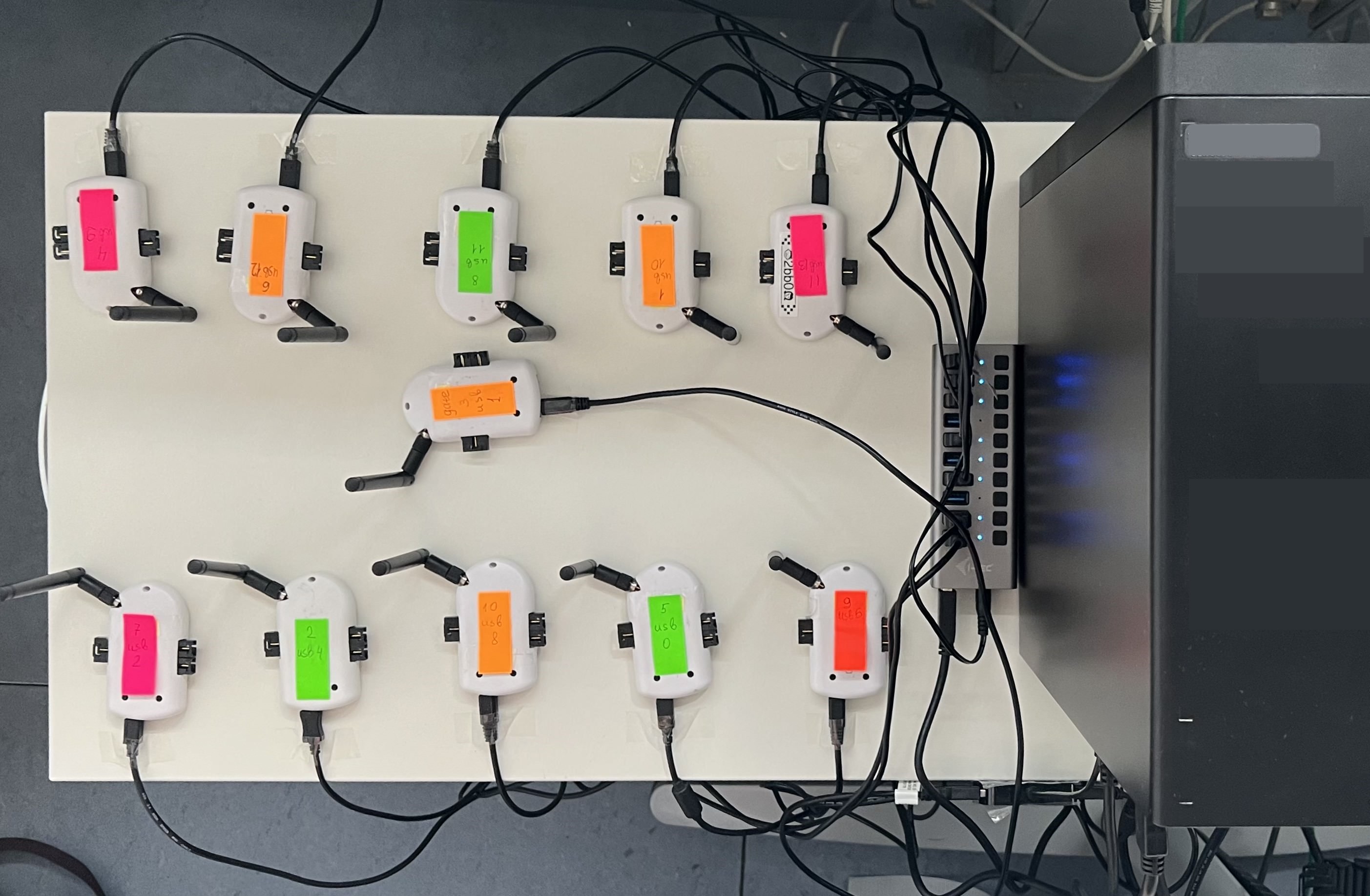}\par
    \label{fig:exp_setup}}
    \vspace*{-2mm}

\caption{The experimental testbed. Plant and controller processes of all loops run in parallel on one PC. Each process is connected to one Zolertia Re-Mote sensor. Sensors communicate over a wireless network.}
\label{fig:testbed}
    \vspace*{-5mm}

\end{figure}

Fig.\ref{fig:testbed} represents the schematic structure and the picture of the measurement setup. The experimental framework consists of $1$ PC and up to $11$ Zolertia nodes, i.e., $5$ control loops and one gateway node. Individual control loops run on the PC as parallel real-time processes implemented in Python. Every $10$ms, the next update is sampled\footnote{Sampling periodicity of $10$ms and the payload size of $20$B, i.e., the traffic intensity, correspond to the closed loop control use case of the 3GPP specification on 5G service requirements for cyber-physical control applications \cite{3gpp.22.104}.}. Sensor processes admit the updates to the network according to the chosen TL policy and push the accepted measurements and the current time step to the node over UART. Sender nodes attempt the transmission, and updates successfully arriving at receiver nodes trigger an ACK.  Receiver nodes use the same MAC and PHY protocols and the same channel for ACKs. Thus, both update messages and ACKs can be delayed or lost. If the sender node receives an ACK, it transfers it to the sensor process on the PC, which stores the statistics of ACKs. Simultaneously, the receiver node sends the state update to the corresponding controller process.

The synchronization between nodes is enforced once at the beginning of each experiment via simultaneous sync command sent by the PC. All the control loops have the same parameters: $n_i = m_i = 1  ~ \forall i$, $A_i = 1.2, B_i = 1, W_i = 1, Q_i=R_i=1 ~\forall i$. The parameters for the TA mechanism are the following. As a heuristic for $f$, we pick $f \triangleq 0.5 stddev ^{\frac{3}{4}}$. A systematic choice of the function $f$ is an interesting research question that is a part of our future work. Other parameters are: $N_b = 10$, $INC(\lambda)\triangleq \lambda + 1$, $DEC(\lambda) \triangleq \lambda/1.1$, $w=100$. These parameters can be adjusted to the particular WNCS scenario.

\section{Experimental Results}
\label{sec:exp_results}

In the following experiments, we compare the LQG-cost performance of different TL schemes. For each considered TL policy and a network setup, we perform $10$ experiment runs, each by $60$s, i.e., $6000$ time steps. We calculate the average LQG cost for each loop for each run as:
\begin{equation}
	    \bar{\mathcal{J}}_i = \dfrac{1}{4001}\sum_{k = 2000}^{6000} \mathcal{J}_i[k],
	    \label{eq:meanAoI}
\end{equation} 
where the first $2000$ time steps are excluded to eliminate the influence of the transient phase at the beginning of the run.

\begin{figure}[t!]
    \includegraphics[width=1\linewidth]{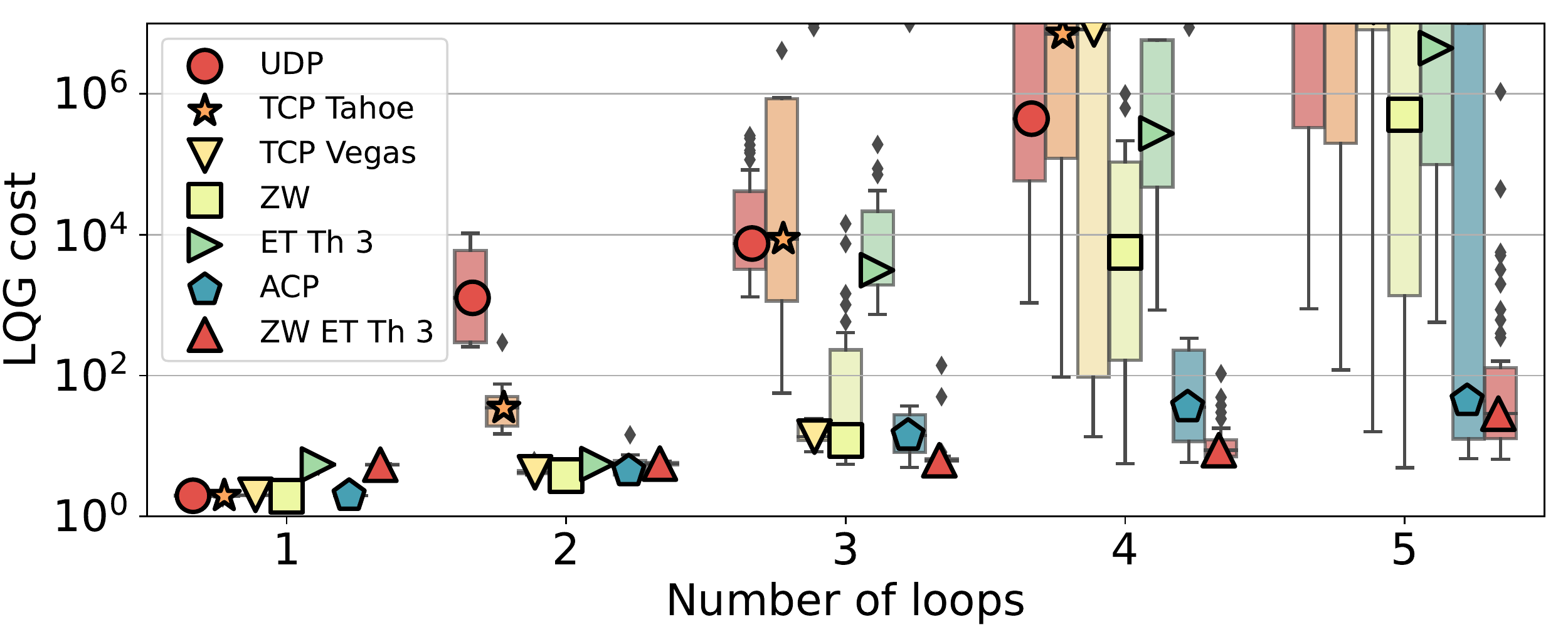}\par

\caption{Control performance of conventional, SotA and ZW ET schemes. Markers denote the medians of corresponding distributions.}
\label{fig:sota}
 \vspace*{-6mm}
\end{figure}

Fig.\ref{fig:sota} depicts LQG cost boxplots for conventional, SotA and proposed ZW ET TL schemes. As expected, UDP has limited control performance when more than two control loops operate simultaneously because of severe network congestion.TCP Tahoe and TCP Vegas show better control performance than UDP because TCP limits network congestion, so updates are not accumulated in MAC buffers as in the case of UDP. However, TCP performance is still poor for a higher number of loops because the congestion control of TCP can admit several updates in a row, occupying network resources. Such periods of resource deficiency for other nodes result in occasional long intervals when the controller does not receive new updates, severely affecting the average control cost. Delay-based TCP Vegas outperforms TCP Tahoe agnostic to networking delays because TCP Vegas limits updates' waiting times. When only one loop is active, UDP and TCP show optimal LQG cost because the end-to-end delay is less than the sampling period of $10$ms, and all the sampled updates are successfully delivered. Congestion-aware ZW policy shows better control performance than TCP due to the eliminated waiting time in the MAC buffers and moderate consumption of resources by individual loops. However, when more than three loops are active, the median LQG cost exceeds $10^3$. The reason for that is that most of the time, the packets of each sensor attempt to access the channel, resulting in increased collision rate and delays. Conventional ET with threshold $\lambda_i=3 ~ \forall i$ shows better control performance than UDP for up to three loops since sensors admit fewer updates to the network. However, as the number of loops increases, the network becomes severely congested, resulting in too performance on LQG cost, because ET, similarly to UDP, is congestion-unaware.

\begin{figure*}[t!]
    \centering
    
    \subfigure[CSMA access scheme]
    {\includegraphics[width=0.49\linewidth]{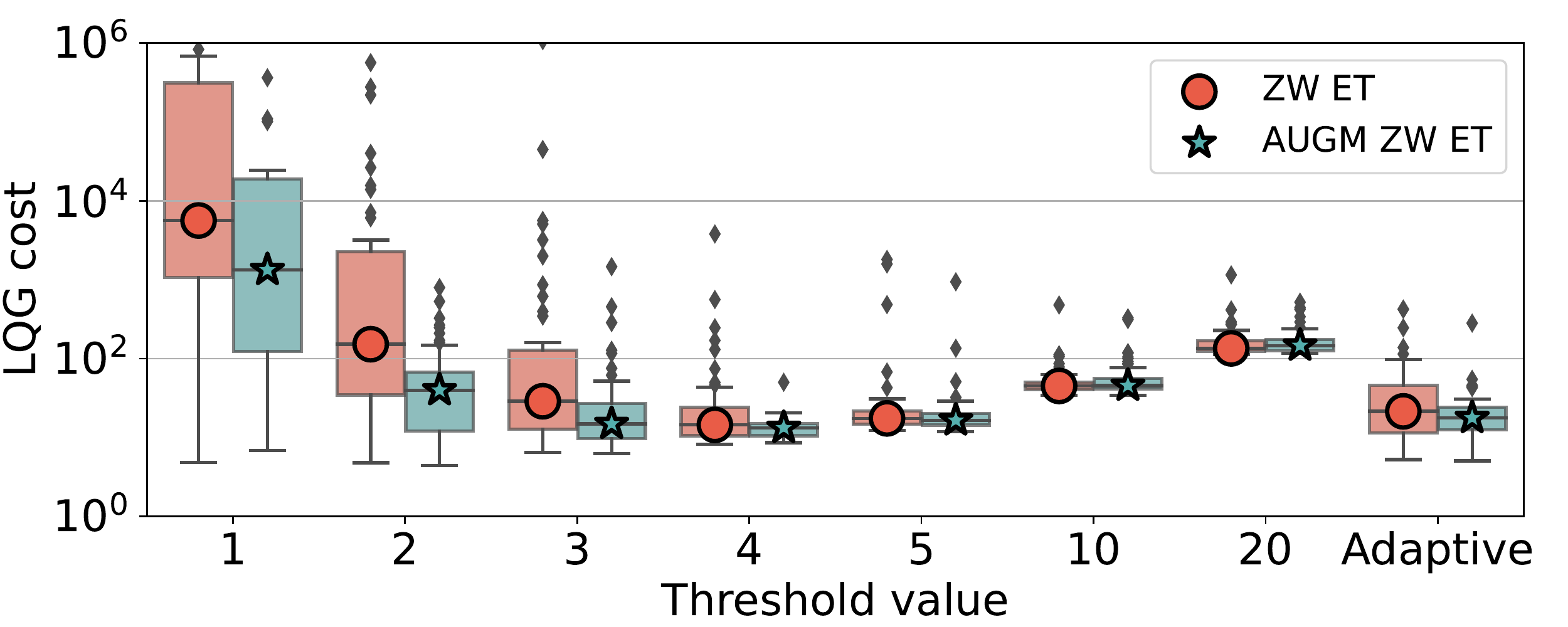}\par
    \label{fig:5_loops_lqg}}
    \subfigure[Polling-based access scheme]{\includegraphics[width=0.49\linewidth]{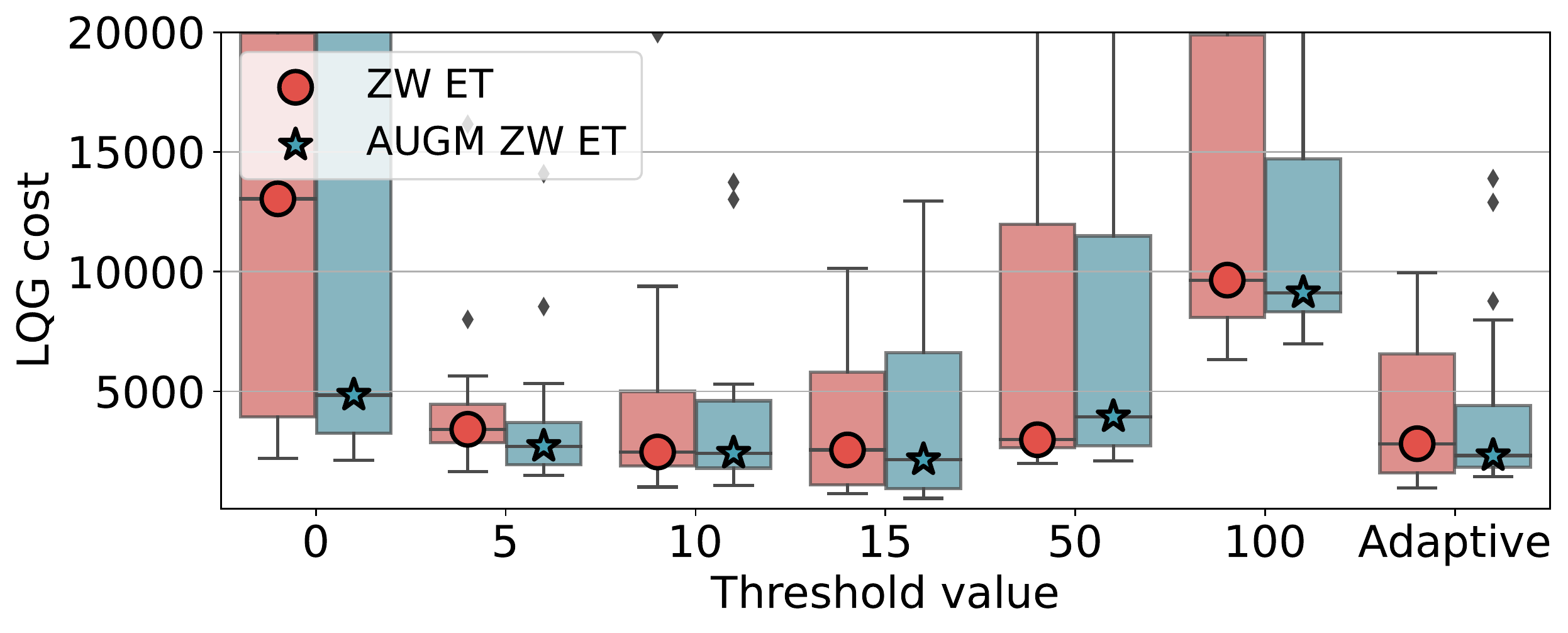}\par
    \label{fig:poll_lqg}}
    \vspace*{-2mm}

\caption{Control performance of proposed ZW ET and AUGM ZW ET schemes for different threshold values and for adaptive threshold. Markers denote the medians of corresponding distributions.}
 \vspace*{-5mm}
\label{fig:exp_lqg}

\end{figure*}

\begin{figure*}[t!]
    \centering
    
    \subfigure[Threshold dynamics]
    {\includegraphics[width=0.49\linewidth]{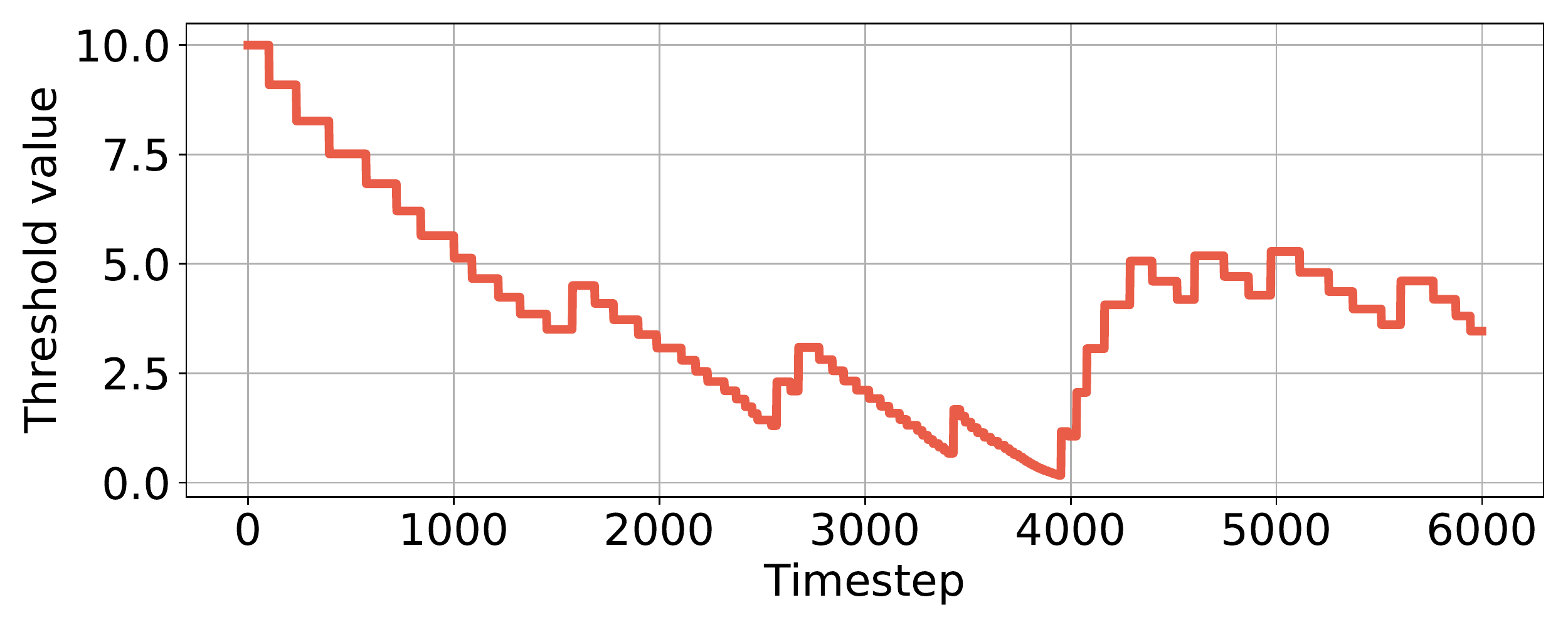}\par
    \label{fig:th_adapt}}
    \subfigure[Control performance]{\includegraphics[width=0.49\linewidth]{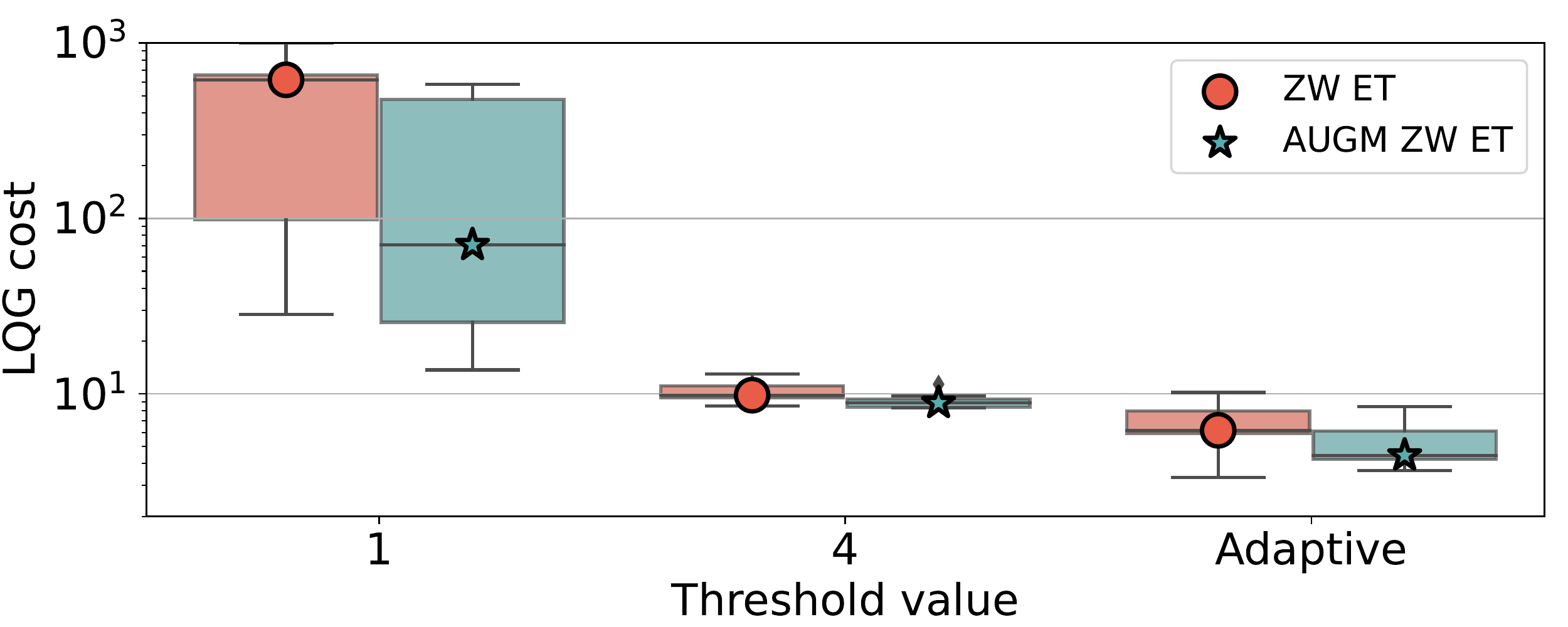}\par
    \label{fig:lqg_adapt}}
    \vspace*{-2mm}

\caption{The performance of the threshold adaptation scheme in the scenario with the dynamic number of loops. One loop is active in the beginning, four more are activated at timestep $4000$. For comparison, the performance with the fixed threshold is given.}
\label{fig:adapt}
    \vspace*{-5mm}

\end{figure*}

The proposed ZW ET scheme outperforms all other considered methods if more than two loops share the network, proving that the consideration of both the relevance of the update for the control process and the current network state allows for achieving better control performance. ZW ET limits the network traffic, allowing for faster delivery of urgent updates when their relevance is detected with admission policy.

Age-minimizing ACP from \cite{shreedhar2019age} has acceptable control performance for $N \le 4$, revealing that AoI is an effective metric for real-time applications. However, the fact that ZW ET outperforms ACP proves that the minimization of age is not equivalent to the minimization of the control cost, and metrics beyond AoI should be considered.

Notably, the performance of ZW ET with $\lambda_i = 3\: \forall i$ is sub-optimal when $N < 3$ because of the too conservative sending rate of ZW ET. In the next experiment, we vary the threshold value. Fig.\ref{fig:5_loops_lqg} presents the LQG cost of ZW ET protocol with $N =5$ for varying thresholds. We can conclude that ZW ET with $\lambda_i = 4\: \forall i$ performs the best, whereas too low and too high thresholds lead to the control performance degradation.

Along with ZW ET, Fig.\ref{fig:5_loops_lqg} shows the results for AUGM ZW ET, where the sensor augments the controller estimation as in \eqref{eq:augm_estimator} and utilizes it for decision-making. Whereas the control performance of AUGM ZW ET is close to ZW ET for threshold values higher than the optimal, there is a substantial improvement for AUGM ZW ET for lower threshold values. The reason is that AUGM ZW ET avoids unnecessary transmissions, which is critical in highly loaded network scenarios. For the optimal threshold value, AUGM ZW ET provides an approximately $10\%$ LQG cost improvement. If a too low threshold is chosen for the given system, the performance gains of AUGM ZW ET proliferate compared to ZW ET.

Next, we analyze the performance of the TA scheme, which is essential when proposed TL techniques are applied in realistic scenarios where the network conditions and the optimal threshold are not known in advance or vary over time. Fig.\ref{fig:5_loops_lqg} shows the performance of proposed TL techniques with the TA mechanism. The TA allows achieving the control performance close to optimal for both ZW ET and AUGM ZW ET schemes. Note that AUGM ZW ET with AT performs at least $10\%$ better than ZW ET with AT because it allows eliminating unnecessary transmissions during periods when network congestion starts to intensify due to the threshold decrease, but the TA mechanism has not captured the congestion yet. 

To prove the versatility of the proposed ET schemes w.r.t. lower layers, we conduct additional experiments with a fundamentally different polling-based MAC scheme. While CSMA is a probabilistic contention scheme,  the polling-based approach represents deterministic scheduling. Fig.\ref{fig:poll_lqg} shows the average LQG cost for $N=5$ for ZW ET and AUGM ZW ET for different threshold values and AT. The LQG cost follows a similar trend for the polling-based access scheme as for CSMA, with the optimal threshold of $\lambda_i \approx 5 \: \forall i$. Similarly, AUGM ZW ET outperforms ZW ET for lower threshold values, and the performance of AT is close to optimal.

In the last experiment, we prove the necessity and superiority of the TA scheme in dynamic network scenarios. For that, we consider the setup where one loop is active at the beginning of the experiment, and in $40$ seconds, $4$ more loops are activated. We compare the performance of the ZW ET and AUGM ZW ET schemes with the proposed TA mechanism and with the optimal thresholds for $1$ or $5$ loops, i.e., $\lambda_i=1$ and $\lambda_i=4 ~ \forall i$. Fig.\ref{fig:th_adapt} gives the example threshold dynamics of the TA mechanism. In the beginning, the threshold decreases to a low value optimal for $1$ loops, and as soon as $4$ more loops are activated, it sharply increases to values close to $4$. Fig.\ref{fig:lqg_adapt} shows the control performance comparison. The TA algorithm achieves the best control performance. The performance with $\lambda_i=1 ~ \forall i$ is poor due to increased congestion. The threshold of $4$, in turn, is too conservative for the initial part of the experiment with one loop. The AT scheme allows achieving around $40\%$ improvement of control performance compared to the fixed threshold of $4$ if with ZW ET. With AUGM ZW ET, we observe further gains of more than $30\%$ compared to ZW ET since AUGM ZW ET prevents unnecessary traffic and control performance degradation during the period when the threshold has not yet adapted to the higher value.

\vspace*{-2mm}
\section{Conclusion}
\vspace*{-1mm}
\label{sec:conclusion}

In the context of network resource management for WNCSs, the schemes considering the relevance of transmitted updates are theoretically shown to be promising for improved performance w.r.t. control cost. In this work, we propose a novel practically feasible transport layer policy design which considers both the relevance of the plant state updates for the control process and the current network conditions. With the help of the testbed with Zolertia Re-Mote sensors used in real industrial IoT applications, we show that our method outperforms considered conventional and SotA policies. With that, we present the first practical implementation framework of relevance- and congestion-aware TL policies. Moreover, we propose a scheme to refine the updates' relevance, which uses the augmentation of the controller estimation at the sensor. In addition, we develop a distributed threshold adaptation scheme tailored for usage in real-life scenarios where network conditions vary for different devices and over time. We stress that the threshold adaptation according to the instantaneous network conditions has not been considered before. The threshold adaptation scheme shows close to optimal performance for scenarios with fixed network conditions and outperforms constant threshold schemes in setups with varying network configurations. Our results prove that relevance- and congestion- awareness are necessary components for TL to achieve better control performance in practical scenarios.

\vspace*{-2mm}
 
\bibliography{main}
\bibliographystyle{IEEEtran}

\end{document}